\documentclass[preprint]{aastex63}

\begin{document}

\shorttitle{Jhelum Stellar Stream}
\shortauthors{Sheffield et al.}


\graphicspath{{figures/}}

\title{Chemodynamically Characterizing the Jhelum Stellar Stream with APOGEE-2}

\correspondingauthor{Allyson A. Sheffield}
\email{asheffield@lagcc.cuny.edu}

\author[0000-0003-2178-8792]{Allyson A. Sheffield}
\affiliation{CUNY - LaGuardia Community College, Department of Natural Sciences, Long Island City, NY 11101, USA}

\author{Aidan Z. Subrahimovic}
\affiliation{CUNY - Macaulay Honors College, 35 W 67th St, New York, NY 10023, USA}
\affiliation{CUNY - City College, Physics Department, 160 Convent Ave, New York, NY 10031, USA}

\author{Mohammad Refat}
\affiliation{CUNY - Baruch College, 55 Lexington Ave, New York, NY 10010, USA}

\author[0000-0002-1691-8217]{Rachael L. Beaton}
\altaffiliation{Hubble Fellow}
\altaffiliation{Carnegie-Princeton Fellow}
\affiliation{Department of Astrophysical Sciences, Princeton University, 4 Ivy Lane, Princeton, NJ~08544, USA}
\affiliation{The Observatories of the Carnegie Institution for Science, 813 Santa Barbara St., Pasadena, CA~91101, USA}

\author{Sten Hasselquist}
\affiliation{Department of Physics and Astronomy, University of Utah, Salt Lake City, UT~84112, USA} 
\affiliation{NSF Astronomy and Astrophysics Postdoctoral Fellow}
\author{Christian R.~Hayes} \affiliation{Department of Astronomy, University of Washington, Seattle, WA, 98195, USA}
\author[0000-0003-0872-7098]{Adrian.~M.~Price-Whelan}
\affiliation{Center for Computational Astrophysics, Flatiron Institute, 162 Fifth Ave, New York, NY 10010, USA}
\author[0000-0003-1856-2151]{Danny Horta}\affiliation{Astrophysics Research Institute, Brownlow Hill,  Liverpool, L3 5RF, UK}

\author{Steven R. Majewski}\affiliation{Department of Astronomy, University of Virginia, Charlottesville, VA 22904-4325, USA}

\author[0000-0001-6476-0576]{Katia Cunha} 
\affiliation{Steward Observatory, The University of Arizona, Tucson, AZ, 85719, USA} \affiliation{Observat\'{o}rio Nacional, 20921-400 So Crist\'{o}vao, Rio de Janeiro, RJ, Brazil}

\author[0000-0002-0134-2024]{Verne V. Smith}\affiliation{National Optical Astronomy Observatories, Tucson, AZ 85719, USA}

\author{Jos\'{e} G. Fern\'{a}ndez-Trincado}
\affiliation{Instituto de Astronom\'ia y Ciencias Planetarias, Universidad de Atacama, Copayapu 485, Copiap\'o, Chile} 
\affiliation{Centro de Investigaci\'on en Astronom\'ia, Universidad Bernardo O Higgins, Avenida Viel 1497, Santiago, Chile}

\author{Jennifer Sobeck}
\affiliation{Department of Astronomy, University of Washington, Seattle, WA, 98195, USA}

\author{Ricardo R. Mu\~{n}oz}
\affiliation{Universidad de Chile, Av. Libertador Bernardo O'Higgins 1058, Santiago de Chile}

\author{D.~A. Garc\'{i}a-Hernández}
\affiliation{Instituto de Astrofísica de Canarias, E-38205 La Laguna, Tenerife, Spain}
\affiliation{Universidad de La Laguna, Departamento de Astrofísica, E-38206 La Laguna, Tenerife, Spain}
\author{Richard R. Lane}\affiliation{Instituto de Astronom\'{i}a y Ciencias Planetarias de Atacama, Universidad de Atacama, Copayapu 485, Copiap\'o, Chile}
\author{Christian Nitschelm}\affiliation{Centro de Astronom\'{i}a, Universidad de Antofagasta, Avenida Angamos 601, Antofagasta 1270300, Chile}
\author{Alexandre Roman-Lopes}\affiliation{Departamento de Física, Facultad de Ciencias, Universidad de La Serena, Cisternas 1200, La Serena, Chile}

\begin{abstract}

We present the kinematic and chemical profiles of red giant stars observed by the APOGEE-2 survey in the direction of the Jhelum stellar stream, a Milky Way substructure located in the inner halo of the Milky Way at a distance from the Sun of $\approx$ 13 kpc. From the six APOGEE-2 Jhelum pointings, we isolate stars with log($g$) $<$ 3.5, leaving a sample of 289 red giant stars. From this sample of APOGEE giants, we identified seven stars that are consistent with the astrometric signal from $Gaia$ DR2 for this stream. Of these seven, one falls onto the RGB along the same sequence as the Jhelum stars presented by \cite{ji20}. This new Jhelum member has [Fe/H]=-2.2 and is at the tip of the red giant branch. By selecting high orbital eccentricity, metal-rich stars, we identify red giants in our APOGEE sample that are likely associated with the $Gaia$-Enceladus-Sausage (GES) merger. We compare the abundance profiles of the Jhelum stars and GES stars and find similar trends in $\alpha$-elements, as expected for low-metallicity populations. However, we find that the orbits for GES and Jhelum stars are not generally consistent with a shared origin. The chemical abundances for the APOGEE Jhelum star and other confirmed members of the stream are similar to stars in known stellar streams and thus are consistent with an accreted dwarf galaxy origin for the progenitor of the stream, although we cannot rule out a globular cluster origin.

\end{abstract}


\section{Introduction}\label{sec:intro}

Stellar streams are relics from the Milky Way's assemblage and thus provide a means of probing the Galaxy's ancient mergers. Large-scale photometric surveys such as 2MASS, SDSS, and PAN-STARRS revealed that the Galactic halo contains a panoply of crisscrossing streams \citep[e.g.,][]{maj03,belo06,slater14}. As photometric surveys probed ever deeper into the Milky Way, a more intricate network of halo streams has emerged, and a spate of new streams have been recently uncovered in the Dark Energy Survey \citep{shipp18} and \emph{Gaia}~DR2 \citep[e.g.,][]{malhan18,ibata19}. To understand fully the nature of individual streams and to explore possible connections between different stellar streams --- some may be tidal remnants of the same merger event --- the full six-dimensional phase space and chemical abundances of its individual members are needed. 

Stellar streams in the halo are remnants from the tidal disruption of either a globular cluster or a dwarf satellite galaxy. In some cases the progenitor is still intact (e.g., as in the cases of the Sagittarius dwarf galaxy and the globular cluster Pal 5) and the streams are still visibly associated with the progenitor, while in other cases both the progenitor and stream may be spatially incoherent and mixed in with the smooth halo (\citealp[e.g.,][]{Schiavon2017,fernandezTrincado19,Horta2021b}). However, chemical abundances (in particular the dispersion in metallicity), mass-to-light ratio, and the ratio of blue horizontal branch to blue straggler stars can all help discern whether the progenitor of a stellar stream is a globular cluster or a dwarf satellite galaxy. Of the eleven new streams reported in \cite{shipp18}, those with relatively thicker stream widths and higher mass-to-light ratios are more likely to be associated with a dwarf galaxy; eight of the new DES streams fall into this category, including the Jhelum stellar stream.  

The Jhelum stream stretches nearly 30$^{\circ}$ across the Southern sky. The morphology of the Jhelum stream is complex, with two spatially distinct parts of the stream identified \citep{bonaca2019} and signs of a gap in the stream. Such gaps in the stream may be indicative of an encounter with a dark matter subhalo, similar to what is seen in the GD-1 stream \citep{grillmair06}. The Jhelum stream's two components also show similar but potentially distinct proper motion signals \citep{malhan18,bonaca2019,shipp19}.
There is also a speculative connection between the Jhelum and Indus stellar streams, based on their orbital properties \citep{bonaca2019}, which could be evidence for multiple wraps of the same stream. Based on its position in phase space ($E-L_{z}$), \cite{bonaca20} find evidence that the Jhelum stream is tidal debris associated with a dwarf galaxy.  

The Southern Stellar Stream Spectroscopy Survey ($S^5$; \citealt{li19}) is designed specifically to explore the kinematics and chemistry of stellar streams in the Milky Way. Medium-resolution spectroscopic data for fourteen Jhelum members are presented in the $S^5$ \citep{li19}. Using follow-up high-resolution MIKE/Magellan spectroscopy, \cite{ji20} present the detailed abundance profiles for the eight brightest stars in the Jhelum stream from the $S^5$ survey. The abundance trends for these eight red giant branch (RGB) stars are consistent with those from disrupted dwarf galaxies.  

Past mergers have played a large role in shaping the Milky Way's outer disk and inner halo. Multiple studies \citep{nissen10,nissen11,hawkins15,hayes18} found evidence for a distinct population in the [$\alpha$/Fe]-[Fe/H] plane consistent with a merger with a massive dwarf galaxy. Using main sequence turnoff stars (MSTO) from SDSS-$Gaia$ DR1, \cite{belokurov18} identified a population of metal-rich stars with nearly no net rotation ($v_{\phi} \approx$ 0 km/s) and a high radial anisotropy is found occupying nearly $\sim$2/3 of the local stellar Galactic halo.  This so-called $Gaia$-Sausage, named for its shape in velocity space, has been traced to a major Milky Way merger 10 Gyr ago. A population of nearby stars comprising a similarly elongated structure in velocity space was found by \cite{helmi18}, using $Gaia$ DR2 and the Apache Point Observatory Galactic Evolution Experiment (APOGEE); these stars have a wider metallicity distribution and are also consistent with a major merger event, dubbed $Gaia$-Enceladus, 10 Gyr ago. In their complementary study, \cite{Haywood_2018} indicated that these structures had an age younger than the bulk of the halo. We will refer to the $Gaia$-Sausage and $Gaia$-Enceladus structures as the GES. The GES event is implicated as a means of heating and puffing up the Milky Way's thick disk.   
Eight globular clusters associated with the GES event have also been detected \citep{myeong18} and are distinct among Milky Way globular clusters. \cite{deason18} found that stars consistent with the GES event show a ``pile up'' at an apogalacticon of around 20 kpc. This finding suggests that the break in the density profile of the halo at this same location could be associated with the GES event. The chemical signature for the GES is explored in APOGEE data \citep{mackereth19}, in particular the ``knee'' at [Fe/H] $\approx$ -1.3 in the [Mg/Fe]-[Fe/H] plane, a value consistent with a massive progenitor. 

In this paper, we present an expanded view of the chemical abundances and kinematics for stars in the Jhelum stellar stream using APOGEE-2 and $Gaia$-DR2. In Section \ref{sec:methods}, we present our astrometric, photometric, and spectroscopic criteria for identifying Jhelum stellar stream giants in APOGEE-2. In Section \ref{sec:kinematics}, we use proper motions from $Gaia$-DR2 to derive the kinematics for the Jhelum stream  as well as stars in the GES. The APOGEE-2 chemical profile for Jhelum is presented in Section \ref{sec:chemistry}, and we discuss our results and their implications in Section \ref{sec:summary}.   

\section{Data and Methodology} \label{sec:methods}

\subsection{APOGEE Data} 

The APOGEE-2 survey \citep{majewski_2017} is one pillar of the Sloan Digital Sky Survey \citep[SDSS-IV;][]{blanton_2017} that makes use of almost-twin infrared spectrographs \citep{wilson_2019} installed at Apache Point Observatory on the 2.5-meter SDSS Foundation Telescope \citep[APO;][]{gunn_2006} and on the Las Campanas Observatory Ir\'en\'ee DuPont Telescope \citep[LCO;][]{Bowen_1973}. 
APOGEE-2 collects $R\sim$22,500 spectra in the $H$-band for stars selected by a set of targeting algorithms described in \citet[][F.~Santana et al.~submitted, AAS29036, R.~Beaton et al.~submitted, AAS29028]{zasowski_2017}. The spectra are reduced and processed through the APOGEE Stellar Parameters and Chemical Abundances pipeline \citep[ASPCAP;][]{garciaperez_2016}. 
This paper makes use of an internal dataset that includes all observations obtained until the cessation of APOGEE observations in March 2020 due to the COVID19 pandemic.
These data are processed identically to the data provided in the earlier SDSS Data Release 16 \citep[DR16;][]{dr16} --- more specifically, with the ASPCAP updates and the calibration adopted for DR16 \citep[][V.~Smith et al.~in prep.]{Jonsson_2020}.

More specifically, the DR16 version of ASPCAP provides the spectra, heliocentric radial velocities, stellar atmosphereic parameters ($T_{\rm eff}$, $\log g$, $v_{micro}$, [$M/H$], [$C/M$], [$N/M$], [$\alpha/M$], $v_{macro)}$) and attempts to measure 26 chemical species (C, C I, N, O, Na, Mg, Al, Si, P, S, K, Ca, Ti, Ti II, V, Cr, Mn, Fe, Co, Ni, Cu, Ge, Rb, Ce, Nd, and Yb). 
What species are successfully measured for a given star depend on $S/N$, its radial velocity, and, to some level, its metallicity. 
Uncertainties are determined based on the fitting uncertainties, the comparison of serendipitous duplicate observations,  and evaluation of abundances within a number of well-sampled star clusters. 
The absolute calibration of the abundances is set to the solar-neighborhood, such that the median abundance for solar-neighborhood stars is shifted to 0 in all abundances for dwarfs and giants separately \citep[see additional details in][]{Jonsson_2020}. 
\citet{Jonsson_2020} provides an element-by-element discussion of the reliability of the DR16 ASPCAP results and we note that many species are less well-measured due to a number of effects (e.g., number of spectral lines, strength of feature, etc.).

\begin{figure}[h]
    \centering
    \includegraphics[width=0.8\textwidth]{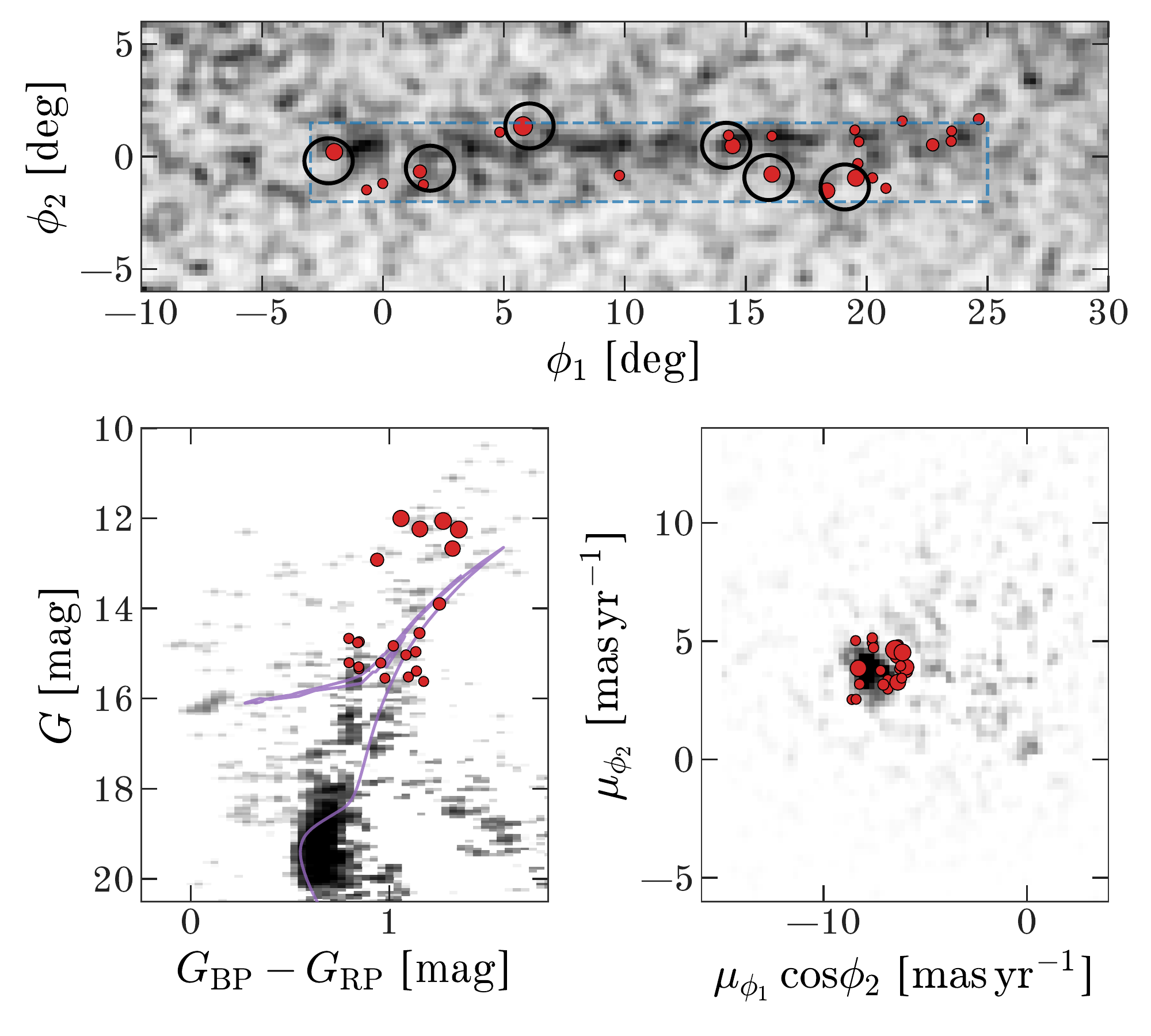} 
    \caption{A summary of the selection criteria
    used to identify RGB candidates of the Jhelum stream to be observed by APOGEE in six fields.
    \textbf{Top:} The background, grayscale image shows the density of MSTO stars
    shown in the Jhelum stream sky coordinates $(\phi_1, \phi_2)$, as defined in \cite{bonaca2019},
    selected using a rectangular box in \textit{Gaia} $G_{\rm BP}-G_{\rm RP}$ color and $G$-band magnitude and a circular selection in proper motion around the mean proper motion of the stream \citep{shipp19}; the MSTO density is only included for context to show where the stream is. 
    The markers (red) in this and all panels show RGB stars selected in \textit{Gaia} color, magnitude, and proper motion as candidate Jhelum stream members.
    The size of the markers indicates the $H$-band magnitude of the sources (larger markers are brighter).
    The circular fields show the six Jhelum stream fields observed by APOGEE, each of which contains one or two of the brightest candidate Jhelum stream RGB stars.
    \textbf{Bottom left:} The \textit{Gaia} color--magnitude diagram of the stream region (outlined with the blue rectangle in the top panel) differenced with the scaled color--magnitude distribution of the stars that do not fall in the stream region (i.e., the stars outside of the blue rectangle in the top panel).
    In both sky regions, we also select stars that have proper motions within $0.5~{\rm mas}~{\rm yr}^{-1}$ of the mean Jhelum stream proper motion \citep{shipp19}.
    The dark overdensity between $18 \lesssim G \lesssim 20$ shows that there is an over-density of MSTO stars: This overdensity is created by Jhelum stream members.
    The isochrone shown is the same as used by \cite{shipp19} and is used only for visualization.
    \textbf{Bottom right:} The proper motions (shown in Jhelum stream coordinates) of stars in the stream region differenced with the scaled proper motion distribution of the other stars in the top panel that do not fall in the stream region. 
    In both regions in the top panel, stars are selected between $18 < G < 20$ and $0.5 < G_{\rm BP}-G_{\rm RP} < 0.9$ to emphasize the stream signal in the differenced distribution.
    The overdensity around $(\mu_{\phi_1}\,\cos\phi_2, \mu_{\phi_2}) \sim (7, 4)~{\rm mas}~{\rm yr}^{-1}$ is due to
    the Jhelum stream.}
    \label{fig:target-select}
\end{figure}

Our study makes particular use of six fields placed to look for Jhelum stream stars in APOGEE-2S (fields with the names ``JHelum1'' through ``JHelum6''; F.~Santana et al.~submitted, AAS29036); the positions of the fields were selected based on the mapping of Jhelum presented in \citet{shipp18} and the local density of targets (see dark circles marking the area of fields 
in the top panel of Figure~\ref{fig:target-select}). The targeting strategy used for the Jhelum fields was specifically tuned to look for candidates using color--magnitude and proper motion selections based on characterizations of the stream in \citet{shipp19} and \citet{bonaca2019}.

The target selection for the six APOGEE-2S Jhelum fields had two phases: (1) selection of Jhelum candidates and (2) an algorithmic selection described in F. Santana et al. submitted, AAS29036); here, we describe the Jhleum candidates. In detail, photometry and astrometry from \emph{Gaia}~DR2 \citep{gaia_overview,gaia_dr2} were first used to select candidate MSTO star members of the stream to reproduce approximately the stars used to characterize the stream in \citet{bonaca2019}; these candidate Jhelum stream MSTO stars were then used to refine the criteria in sky position, color and magnitude, and proper motions to identify candidate red giant branch (RGB) star members of the stream. Figure~\ref{fig:target-select} shows the Jhelum stream in sky position (top panel; in stream coordinates $\phi_1, \phi_2$), \textit{Gaia} color--magnitude space (bottom left panel), and proper motions (bottom right panel; also in Jhelum stream coordinates), shown as the grayscale image in the background of each panel.
Motivated by the distribution of the MSTO stars in these quantities, RGB stars were selected to (1) lie in the over-dense sky region delineated by the rectangular field indicated in the top panel of Figure~\ref{fig:target-select}, (2) have $G_{\rm BP}-G_{\rm RP} > 0.8$ and 2MASS $H < 14$ (using the official \textit{Gaia}--2MASS crossmatch),
and (3) have proper motions consistent with the mean Jhelum stream MSTO proper motion (these APOGEE criteria for star selection in the Jhelum stream fields are described in more detail in Santana et al.~submitted, AAS29036).
In more detail, to define our proper motion criteria, we construct a model for the non-stream region (i.e., outside of the rectangular window in the top panel of Figure~\ref{fig:target-select}) proper motion distribution, then model the proper motion distribution in the stream region as a mixture of the ``background'' model and an additional Gaussian component meant to represent the Jhelum stream proper motion distribution.
This proper motion model allows us to compute kinematic stream membership probabilities for all RGB stars in this sky region, which we use (combined with the color--magnitude selection) to identify candidate stream RGB stars.
The brightest stars from this selection were then used to define the six Jhelum field positions (dark circles in the top panel of Figure~\ref{fig:target-select}).

With a field-diameter of of 1.5$^{\circ}$, the yield of red giant candidates per APOGEE-2S field was small; candidates were selected using this schema will have \texttt{APOGEE2\_TARGET1} bit 18 set in the APOGEE database. 
The remainder of the targets for each field were selected using the procedures for the altered halo targeting strategy (described in detail by F.~Santana et al.~submitted, AAS29036) that used proper motion priors to remove contamination from nearby dwarfs; the normal APOGEE magnitude limits were relaxed to $H\sim12$ to fill plates with giant candidates.
Much of the scientific analysis that follows is ``blind'' to this input strategy.

The six Jhelum pointings amount to 1583 stars of all stellar types; 80\% of these stars have $S/N > 70$ and 54\% of these have $S/N>100$. 
The target $S/N$ for ``best'' ASPCAP performance is 100, but the uncertainties do not seem to change substantially between $70 < S/N < 100$ \citep{Jonsson_2020}.
Among the Jhelum sample, 374 stars have giant-like $\log g$ and 336 of these (90\%) have $S/N > 70$. 
For the giants, the median uncertainties across all abundances are: $\sigma_{Teff}=100$ K, $\sigma_{logg}=0.06$ dex, $\sigma_{M/H}=0.01$, $\sigma_{\alpha/H}=0.01$, $\sigma_{Mg/Fe}=0.01$, $\sigma_{Ca/Fe}=0.02$, $\sigma_{Si/Fe}=0.02$, and $\sigma_{O/Fe}=0.01$.
However, we note that uncertainties for metal poor stars, as are expected in the Jhelum stream, will be larger due to weaker line expression (typical $\sim$2-3$\times$, but it depends on the element).

\subsection{Astrometry}
\begin{figure*}
  \begin{minipage}[l]{0.5\textwidth}
    \includegraphics[width= 0.9\linewidth]{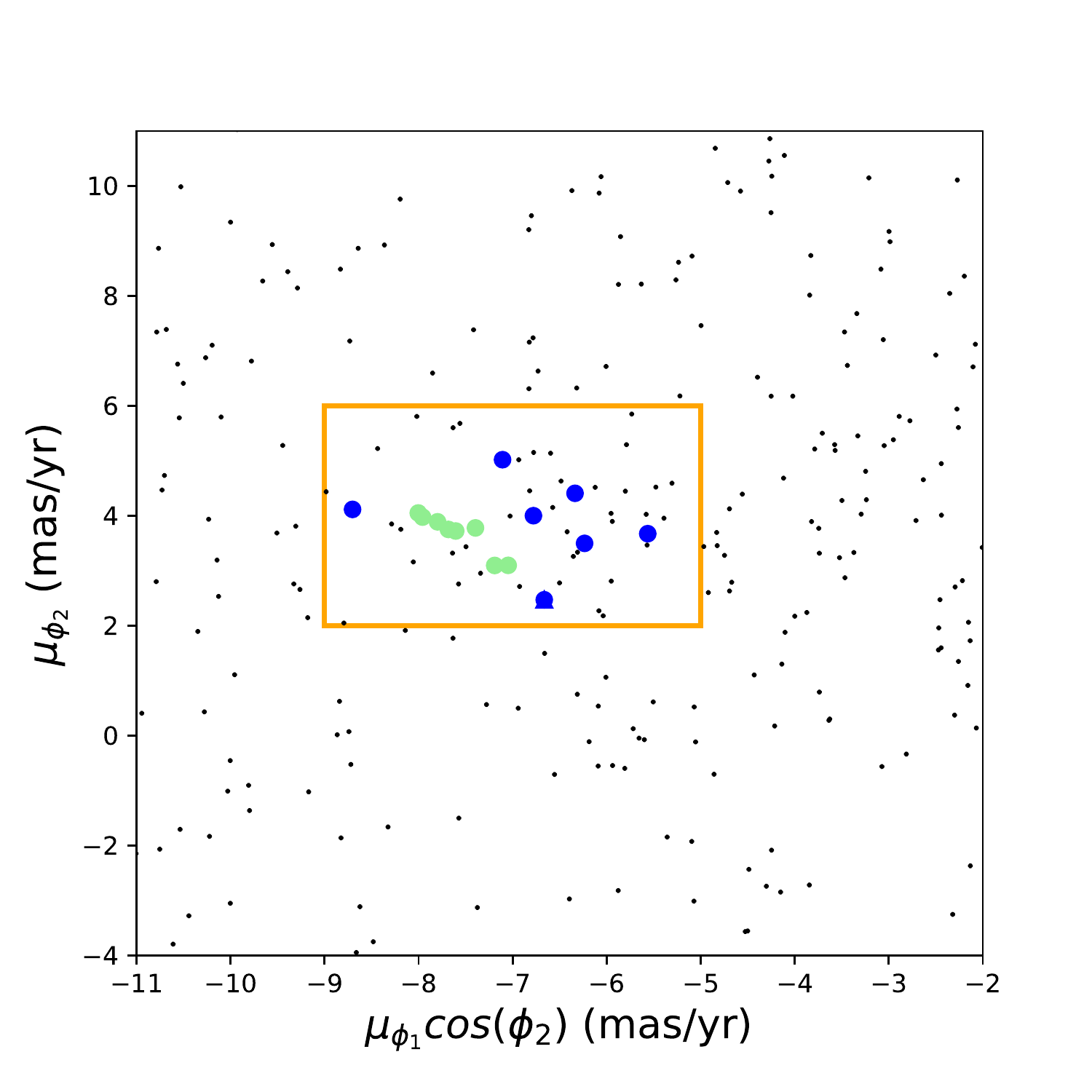} \\
    \includegraphics[width= 0.9\linewidth]{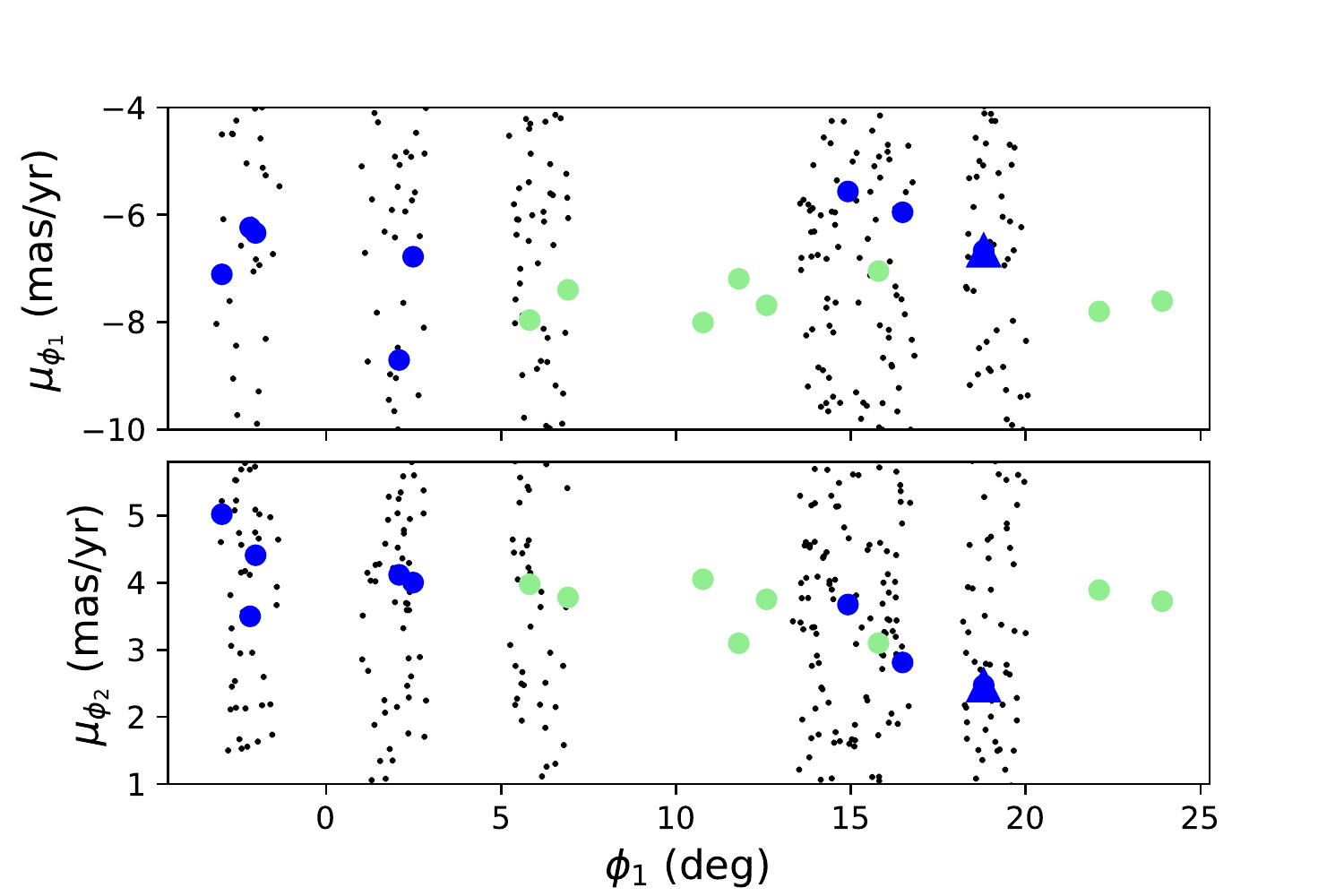}
  \end{minipage}%
  \begin{minipage}[r]{0.5\textwidth}
    \includegraphics[width= 0.9\linewidth,height=10cm]{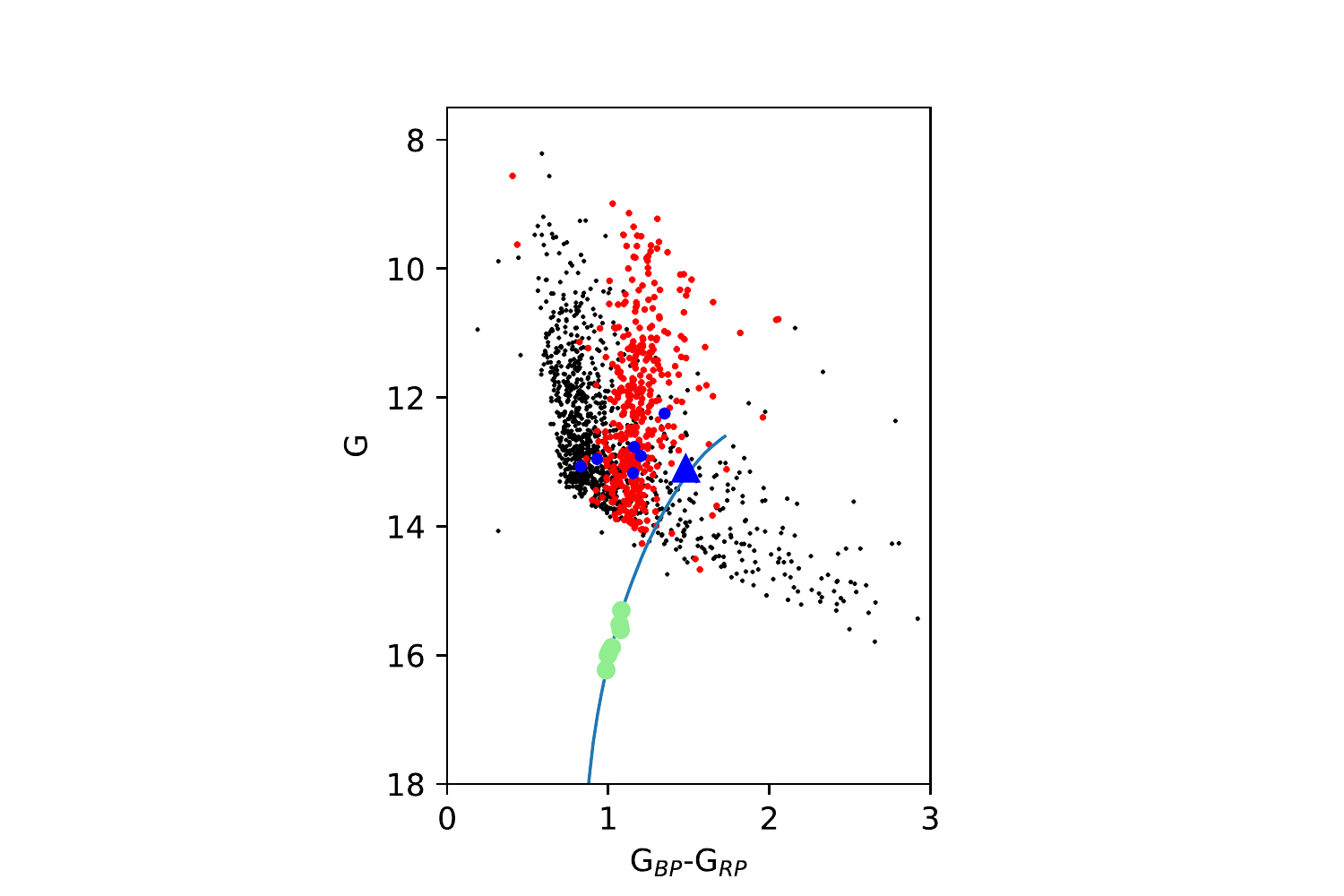}
  \end{minipage}
  \caption{Astrometry and photometry for the APOGEE Jhelum pointing stars, APOGEE Jhelum candidates, and Jhelum stars from \cite{ji20}. In all panels, the APOGEE Jhelum pointing stars are shown as the small black points, the APOGEE giants falling within the proper motion constraints are shown in blue, and the Jhelum stars from \cite{ji20} are shown in light green. The blue triangle is the APOGEE Jhelum giant.
  \textbf{Top left:} The proper motions along the Jhelum stream. The orange rectangle outlines the proper motion region used to isolate Jhelum stream members, and stars with [Fe/H]$<$-1.1 are marked in blue.
  \textbf{Bottom left:} Proper motions as a function of stellar position
  along the Jhelum stream, where the top shows the proper motion along the stream \textbf{($\mu_{\phi_{1}}$)} and the bottom shows the proper motion perpendicular to the stream \textbf{($\mu_{\phi_{2}}$)}. The candidate APOGEE Jhelum giants (blue points) follow a similar trend as the \cite{ji20} Jhelum giants.  
  \textbf{Right:} $Gaia$ color-magnitude diagram. The stars shown in red are selected to be giants and have APOGEE-derived $log(g)<3.5$. The best-fit isochrone \citep{dotter08}, shown as the thin blue line, is for a 12 Gyr population with [Fe/H] $=-1.4$ and [$\alpha$/Fe] $=+0.4$.}
  \label{fig:phot}
\end{figure*}

The location on the sky of the six chosen APOGEE Jhelum pointings is shown in the top panel of Figure \ref{fig:target-select}. The $Gaia$ DR2 proper motions for all of the Jhelum pointing stars are shown in Figure \ref{fig:phot} (black points). Here, we include the proper motions for the known Jhelum stars from \cite{ji20} (light green points). A box, with boundaries -9 $< \mu_{\phi_{1}}^{*} < $ -5 mas yr$^{-1}$, 2 $ <  \mu_{\phi_{2}} < $ 6 mas yr$^{-1}$, is drawn to show our proper motion selection criterion in Figure \ref{fig:phot}. Our candidate Jhelum stars shown in blue. These stars, in addition to falling into the proper motion box, have been selected to have [Fe/H] $<$ $-1.1$, based on isochrone fits \citep{malhan18,bonaca2019} which suggest that the Jhelum stream stars are metal-poor ([Fe/H] $\lesssim$ $-1.5$). 

We show the proper motions along the stream ($\phi_{1}$) in Figure \ref{fig:phot}, with the Jhelum giant stars from \cite{ji20} included for comparison. The stars fall along trends similar to those found by \cite{bonaca2019}, although with a slightly higher dispersion in both proper motion coordinates. One of the APOGEE stars selected to be in the proper motion box shown in Figure \ref{fig:phot} falls along the best-fit isochrone for Jhelum, as shown in the next section. 

\subsection{Photometry}
The first detection of the Jhelum stellar stream was found in the Dark Energy Survey by \cite{shipp18}. A matched-filter search in color-magnitude space was used by \cite{shipp18} to trace out the main sequence for the Jhelum stellar stream. The best-fit isochrone (Dartmouth Stellar Evolution Database; \citealt{dotter08}) for the Jhelum stream from the matching technique by \cite{shipp18} was found to have [Fe/H] = $-1.2$ and age = 12.0 Gyr. 

The Jhelum pointing stars that fall into our proper motion selection box in Figure \ref{fig:phot} in this study span $G_{0}$ $\approx$ 12 to 13, as the APOGEE-2 targeting is only accessing the stars at the tip of the red giant branch (see Figure \ref{fig:target-select}). To select RGB stars from the Jhelum pointing sample, we make a cut on the ASPCAP-derived gravity values of log($g$) $<$ 3.5. This reduces the sample size from 1495 to 289 stars.  

Figure \ref{fig:phot} shows the color-magnitude diagram for all 1495 Jhelum pointing stars, with Gaia $G$, $G_{BP}$, and $G_{RP}$ magnitudes. The Gaia magnitudes have been corrected for extinction using the dust maps of \cite{schlegel98} and using the extinction laws reported in \cite{malhan18}. The stars that fall outside of this $\log{g} < 3.5$ cut are shown as the black points, while the RGB stars are shown in red, and the stars in blue have [Fe/H] $<$ $-1.1$ and fall within the proper motion box. The Jhelum red giant stars from \citet{ji20} fall along the lower RGB track. For this study, we find that the best fit isochrone for the Jhelum RGB stars, using the Dartmouth Stellar Evolution Database \cite{dotter08}, is for a 12 Gyr old population with [Fe/H] = $-1.4$ and [$\alpha$/Fe] = 0.4. We note here that isochrone fitting was not meant to constrain the age or metallicity of the stream, but rather to show that the APOGEE Jhelum giant and the \cite{ji20} Jhelum giants belong to the same stellar population. Based on the CMD, we find one APOGEE giant star from the proper motion box shown in Figure \ref{fig:phot} that falls along the Jhelum RGB trend; we mark this star as the blue triangle. We will refer to this one star as the APOGEE Jhelum giant.

\subsection{Radial Velocities}
\begin{figure}
\begin{center}
    \includegraphics{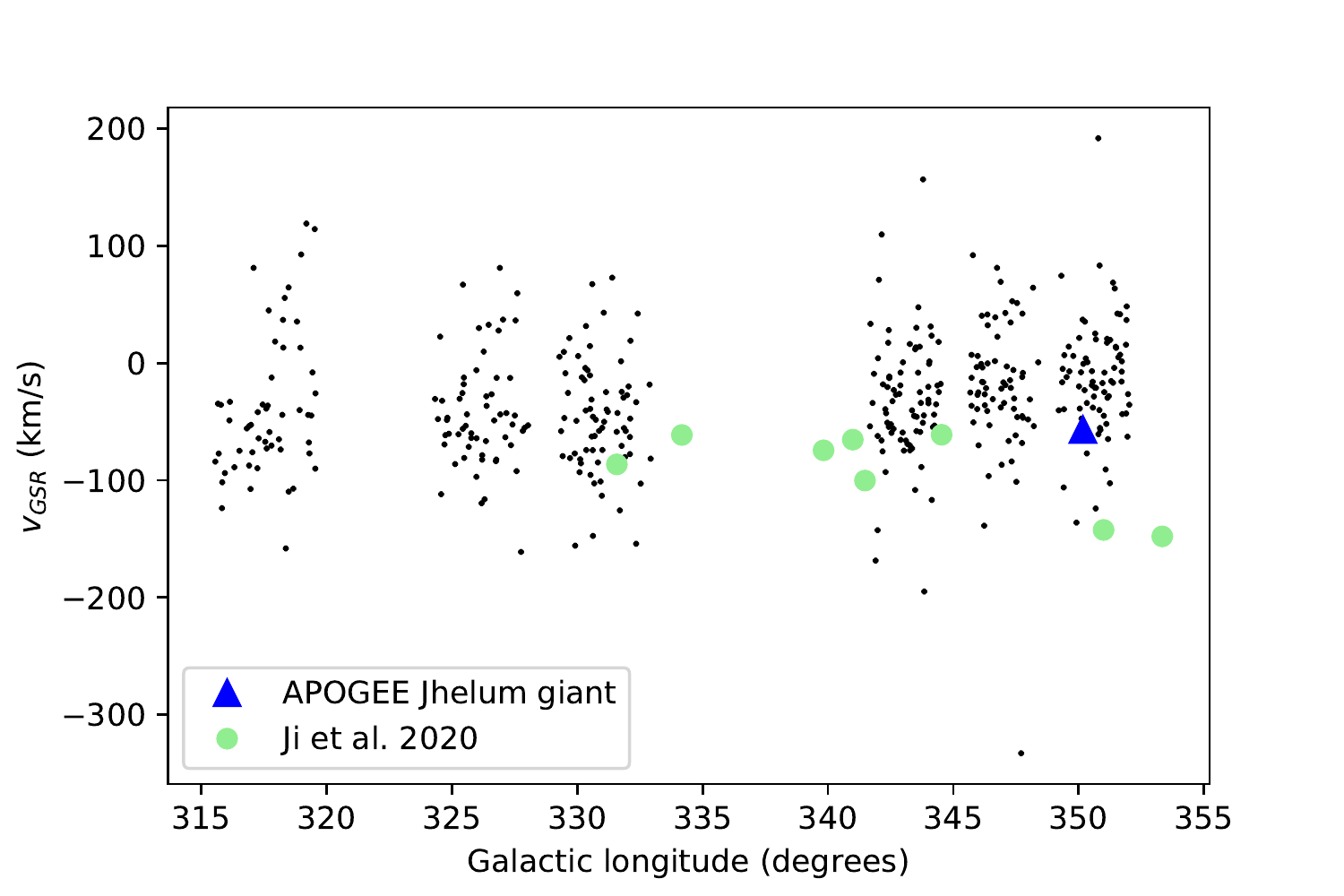}
\caption{Radial velocities in the GSR frame as a function of Galactic longitude. Shown are the Jhelum stars from \cite{ji20} (green points) and the APOGEE-2 Jhelum giant (blue triangle) that falls into the proper motion box shown in Figure \ref{fig:phot} and along the RGB in Figure \ref{fig:phot}.}
\label{fig:vgsr_l}
\end{center}
\end{figure}

The radial velocities in the Galactic standard of rest (GSR) frame\footnote{We adopt $(u,v,w)$=(-11.1, 242.0, 7.25) km/s for the solar motion \citep{schonrich10,bovy12}.} for the APOGEE Jhelum pointing stars selected to be red giant stars are shown in Figure \ref{fig:vgsr_l}. The APOGEE Jhelum giant is shown in blue and the Jhelum red giants from \cite{ji20} are also shown. 

The Jhelum giant stars in Figure \ref{fig:vgsr_l} collectively follow a similar trend in their radial velocities. Two stars from \citet{ji20}, Jhelum2$\_$14 and Jhelum2$\_$15, have lower radial velocities than either the other six Jhelum giants shown in light green or the APOGEE Jhelum giant. The radial velocity profile as a function of declination for the Jhelum stream is seen to be very steep \citep[see Figure 10 in][]{li19}, and the APOGEE Jhelum giant falls along the trend, which also includes all eight Jhelum giants from \citet{ji20}.

In Table 1, we present the position, $Gaia$ DR2 proper motions and magnitudes, and APOGEE-derived radial velocity for the APOGEE Jhelum giant. 

\begin{table}
\footnotesize
\begin{center}
\begin{tabular}{lllllllll}
\hline
APOGEE ID &RA &        Dec &     $G$ &    $G_{BP}$ &    $G_{RP}$ &      $\mu_{\alpha}$ &     $\mu_{\delta}$ &       $v_{hel}$  \\
&(degrees) &        (degrees) &     mag &    mag &    mag &      (mas/yr) &     (mas/yr) &       (km/s)  \\
\hline
2M22033201-4705352 &  330.883394 & -47.093113 &  13.10 &  13.79 &  12.31 &  5.26 $\pm$ 0.023 & -4.78 $\pm$ 0.030 & -32.41 $\pm$ 0.06  \\
\hline
\end{tabular}
\caption{The photometric magnitudes and proper motions are from $Gaia$-DR2 and the radial velocity is from APOGEE.}
\end{center}
\end{table}

\section{Kinematics} \label{sec:kinematics}

Kinematics for the APOGEE-2 Jhelum pointing stars are estimated using the \emph{Gaia}~DR2 \citep{gaia_overview,gaia_dr2} proper motions, APOGEE radial velocities, and distances from the APOGEE value-added $astroNN$ catalog\footnote{\url{https://www.sdss.org/dr16/data_access/value-added-catalogs/?vac_id=the-astronn-catalog-of-abundances,-distances,-and-ages-for-apogee-dr16-stars}} \citep{leung19a,leung19b}. As noted in \cite{bovy19}, the $astroNN$ distance are underestimated for stars with $d >$ 4 kpc. We use the prescription described in \cite{bovy19} to correct the distances for all of the Jhelum pointing stars; the $Gaia$ parallax errors are too large relative to the parallax signal to use for this study. The APOGEE-2 Jhelum giant has a $astroNN$ distance of $12.6 \pm 2.0$ kpc. A distance of $13.2 \pm 2.5$ kpc is adopted for the \citet{ji20} Jhelum giants \citep{shipp18}. The errors on the derived kinematic parameters are correspondingly large, up to 100 km/s for the cylindrical velocities. We note that the errors for the APOGEE Jhelum giant are anomalously high relative to the other APOGEE giants presented in this work, with a typical error at the level of 5$\%$.

\subsection{Gaia-Enceladus-Sausage stars} \label{subsec:sausage}
\begin{figure}
    \centering
    \includegraphics[width=\textwidth]{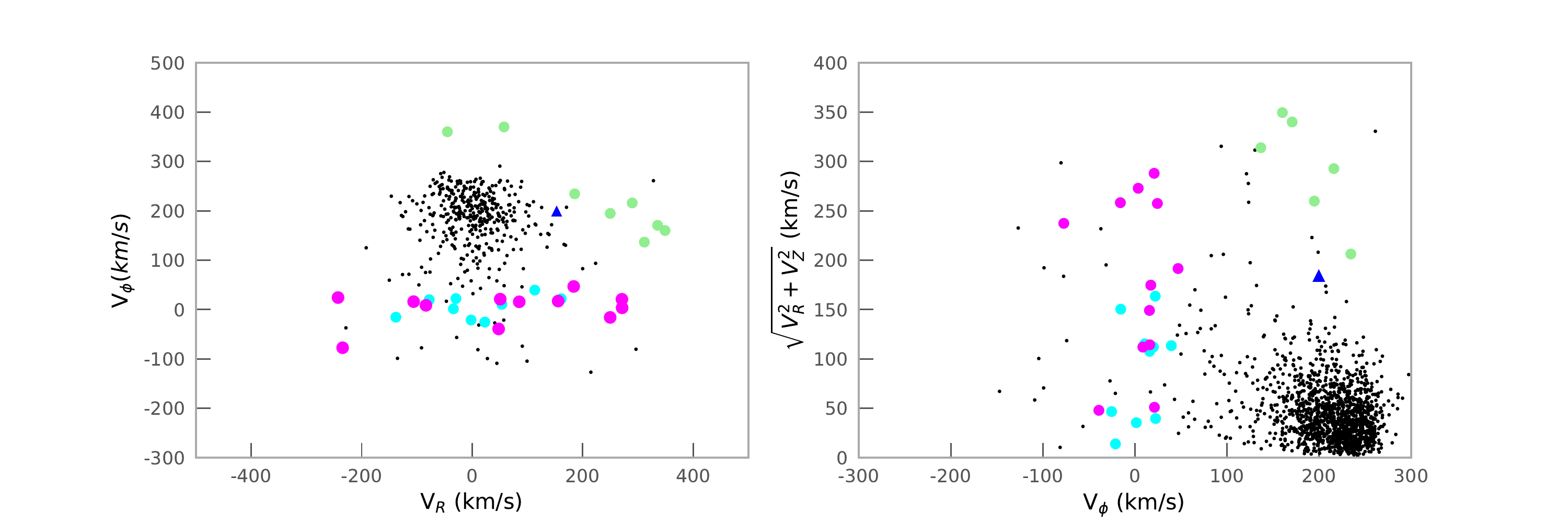}
    \caption{Cylindrical velocities for APOGEE giants in the Jhelum pointing fields (black points). On the left, the azimuthal and radial components of the motion are shown, with the APOGEE Jhelum giant (blue triangle), and the Jhelum giants from \cite{ji20} (light green points), \textbf{with an estimated distance of 13.2 kpc}. APOGEE giants selected to be metal-rich ([Fe/H]$>$-1.7) and have high eccentricities ($e>0.8$) are shown in cyan and magenta, where the cyan points have high-Al ([Al/Fe]$>$0) and the magenta points have low-Al ([Al/Fe]$<$0) abundances derived from ASPCAP. The high-metallicity, high-$e$ stars have motions that are very similar to those from the GES merger debris. On the right, these stars are shown in a Toomre diagram. The Jhelum giants occupy a region that is distinct from the locus that the candidate GES merger stars (cyan and magenta).}
    \label{fig:vcyl}
\end{figure}

To assess the possibility of contamination by GES stars in our sample, we isolate metal-rich giants with high eccentricities, consistent with the detection from \cite{belokurov18}. In Figure~\ref{fig:vcyl}, we show the azimuthal and radial motions in the left panel and the Toomre diagram on the right. In Figure~\ref{fig:vcyl}, stars with $e> 0.8$ and [Fe/H] $>$ -1.7 \citep[e.g.,][]{mackereth19} are shown in cyan and magenta, where the cyan points have [Al/Fe]$>$0 and the magenta points have [Al/Fe]$<$0. 

In the azimuthal-radial motion plane on the right, these high-$e$ metal-rich stars all fall into the region where GES stars were detected by \cite{belokurov18}. Those high-$e$, metal-rich stars with low Al (shown in magenta) are considered to be the most likely GES members, as these low Al values are a reasonable assumption for an accreted star at this metallicity \citep[e.g.,][]{horta21}.  

In the Toomre diagram on the right, the likely GES stars (magenta) occupy the higher energy orbits, with a mix of prograde and retrograde orbits. The APOGEE Jhelum giant, however, falls into the region occupied primarily by prograde disk stars. In the following sections, we compare our APOGEE Jhelum giant with the APOGEE GES stars in our study.

\subsection{Orbits} \label{subsubsec:orbits}

\begin{figure}[h]
    \centering
    \includegraphics[width=0.8\textwidth]{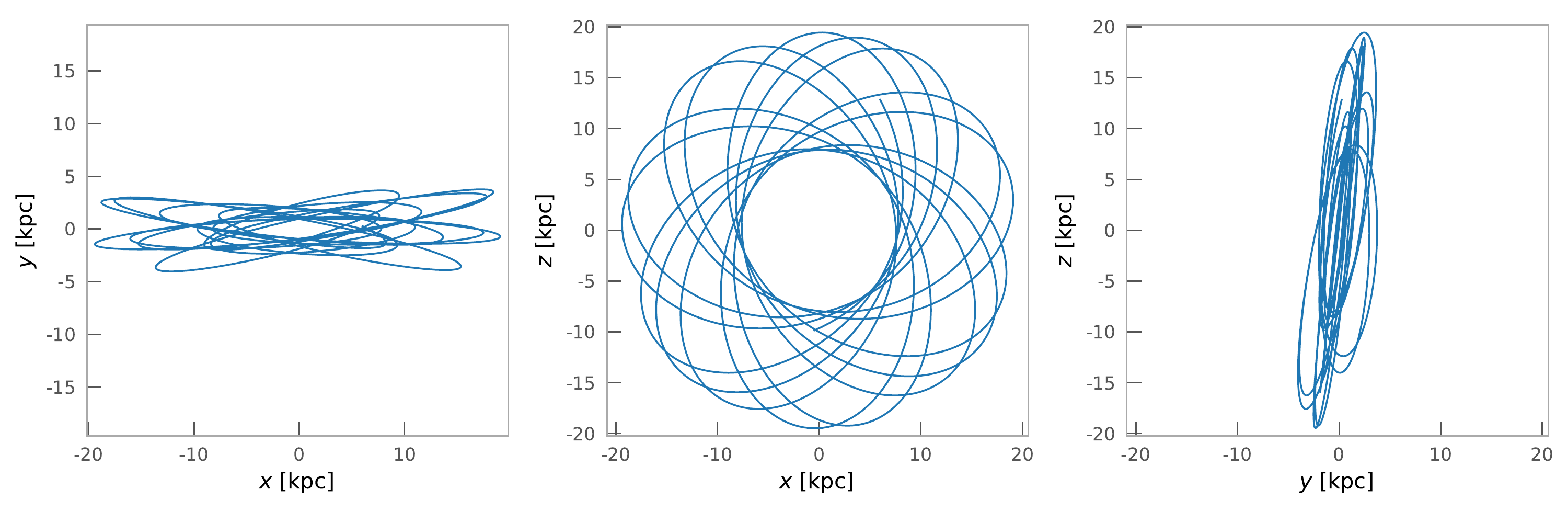} \\
    \includegraphics[width=0.8\textwidth]{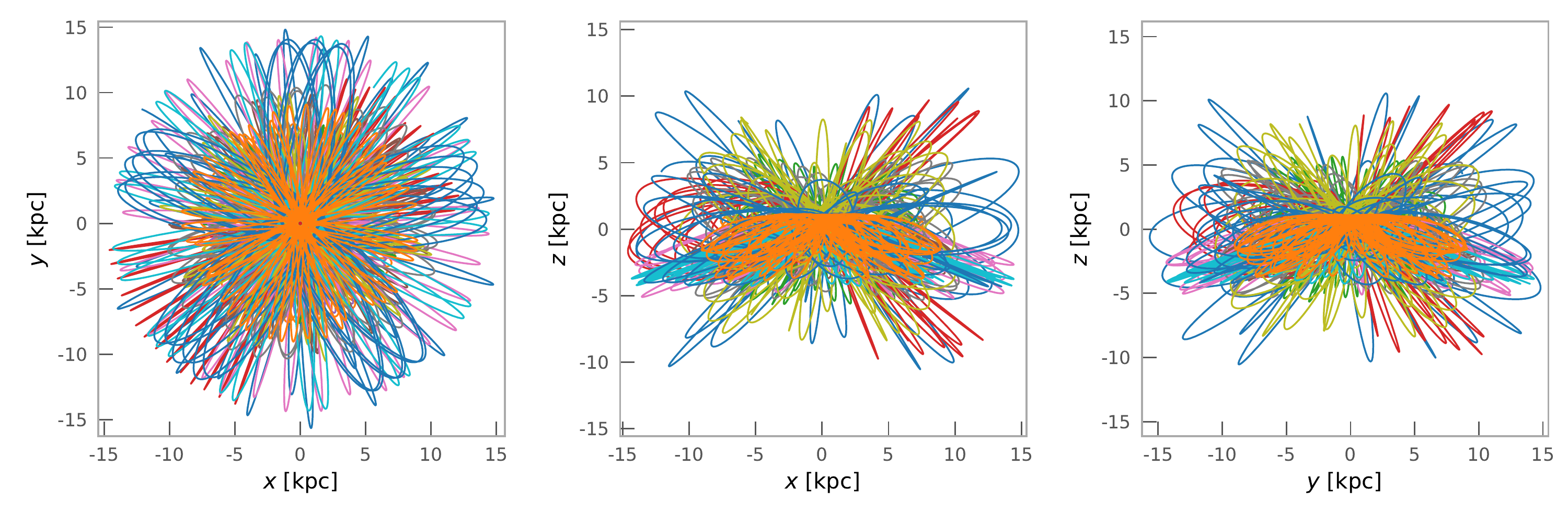}
    \caption{Orbits for the APOGEE Jhelum giant (top) and potential APOGEE GES stars (bottom). The stars shown in the bottom panel (12 total) were selected to have $e>0.8$, [Fe/H] $>-1.7$, and [Al/Fe] $<$ 0.}
    \label{fig:orbits}
\end{figure}

To visualize the orbits of the Jhelum stream giants, we use the orbit integration function in the Python \texttt{gala} package \citep{gala, gala:zenodo}. For these orbits, we use the default \texttt{MilkyWayPotential} as our mass model, which uses a Hernquist potential for the galactic nucleus and bulge, a Miyamoto-Nagai potential for disk stars \citep{2015ApJS..216...29B}, and a spherical NFW profile for stars in the halo of the Milky Way. We use a Leapfrog integration scheme to compute the orbits. The integrator takes four parameters to determine the orbit of an object: distance, equatorial coordinates, proper motions, and radial velocity. Figure \ref{fig:orbits} depicts the projections of these orbits onto the three Cartesian planes integrated over a period of 5 Gyr with a time-step of 0.5~Myr.

\begin{figure}[h]
    \centering
    \includegraphics[width=0.8\textwidth]{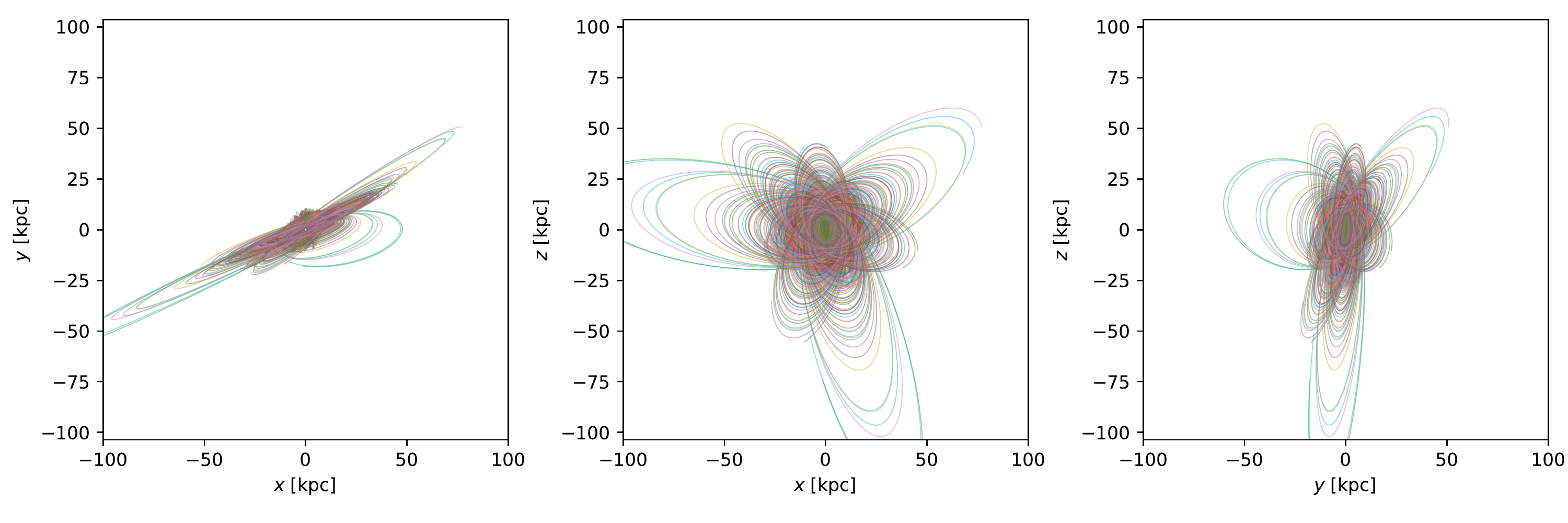} \\
    \includegraphics[width=0.8\textwidth]{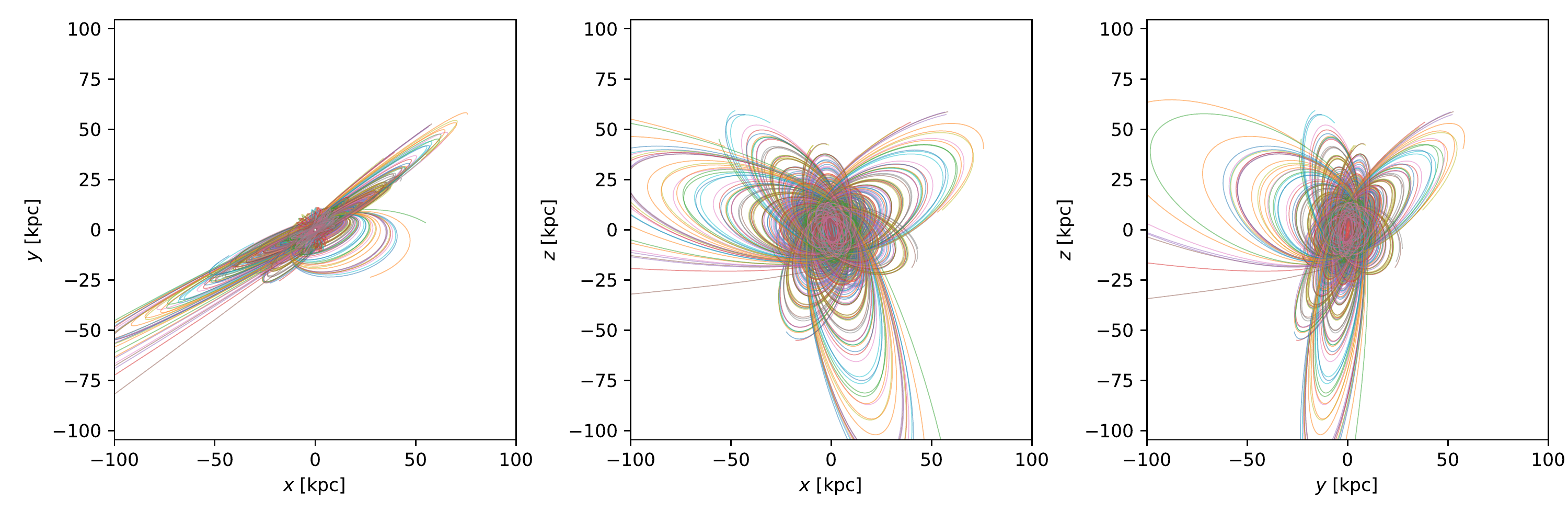}
    \caption{Orbits for (top) the Jhelum red giant and (bottom) one of the Jhelum giants from \cite{ji20} (Jhelum1$\_$15 - $Gaia$ ID 6511949016704646272).}
    \label{fig:orbits_dist}
\end{figure}

In Figure \ref{fig:orbits}, we show orbits for our APOGEE Jhelum giant (top panel) and high-eccentricity metal-rich stars with [Al/Fe] $<$ 0 that fall into the GES region in Figure \ref{fig:vcyl} (bottom panel). The orbits of the APOGEE GES stars compared with the APOGEE Jhelum giant seem inconsistent with stars originating from the same population. Although we do not have well-determined distances for the \cite{ji20} Jhelum giants, an estimate of the general shape of the orbits can be gleaned by sampling the error distribution for the distances. To do this, the full covariance matrix from $Gaia$ was used, which allows us to generate error samples over sky position and proper motion that take into account the reported correlations between these quantities. These error samples were used to generate 128 distance and radial velocity samples drawn from a Gaussian distribution. A comparison for the APOGEE Jhelum giant and one of the Jhelum giants from \cite{ji20} is shown in Figure \ref{fig:orbits_dist} and we note that the orbits for all eight of the Jhelum giants from \citet{ji20} are very similar. It is seen in Figure \ref{fig:orbits_dist} that the orbits our APOGEE Jhelum giant and the Jhelum giants from \cite{ji20} are consistent with stars that originated from the same population.

\subsection{Phase Space} \label{subsubsec:phase}
Because energy and angular momentum are conserved quantities, debris from an accretion event, such as a tidally disrupted cluster or satellite, should tend to cluster in phase space. Trends in $E-L_{Z}$ for GCs and dwarf galaxy stars have been seen recently for APOGEE stars by \cite{horta20} and \cite{horta21}, as well as for stars with $|b|>$ 40$^\circ$ from the H3 Spectroscopic Survey \citep{naidu20} and $Gaia$-DR3 \citep{bonaca20}. The GES debris occupies a fairly narrow locus in $L_{Z}$ centered at 0, as expected from the fact that the GES has nearly no net rotation. Other Milky Way substructures, such as the Helmi streams \citep{helmi99}, occupy distinct regions as well. 

\begin{figure}
\begin{center}
    \includegraphics[width=0.8\textwidth]{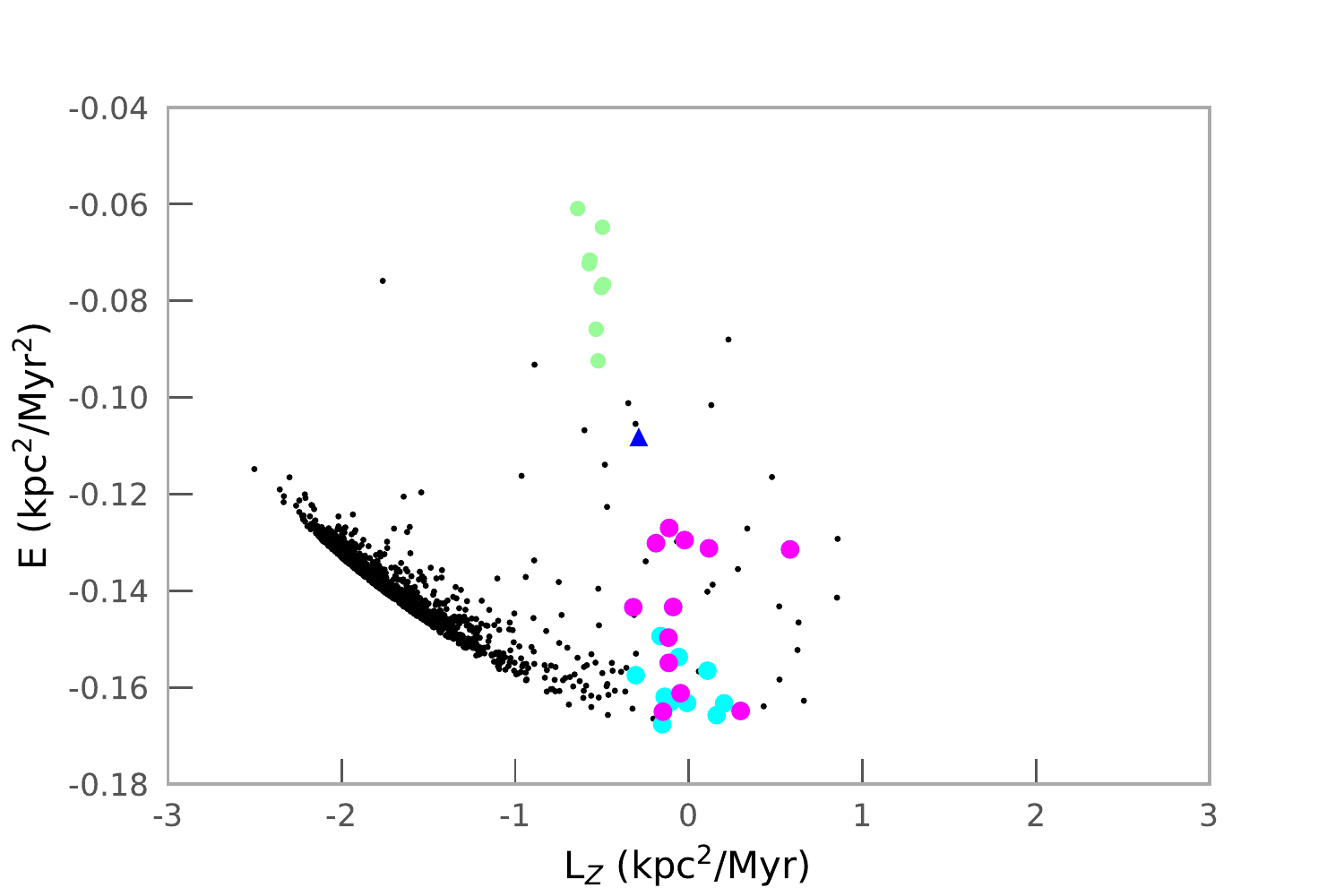}
\caption{Orbital energy as a function of angular momentum for the APOGEE Jhelum pointing stars, with the same color scheme as in Figure \ref{fig:vcyl}. The Jhelum giants occupy a narrow prograde locus, while the APOGEE GES candidates have lower energies and have both prograde and retrograde motions. The stars shown in cyan (metal-rich, high $e$, low-Al) generally have the lowest orbital energies and are consistent with stars from the ``Splash''.}
\label{fig:elz}
\end{center}
\end{figure}

In Figure \ref{fig:elz}, we show the $E-L_{Z}$ distribution for all of the Jhelum pointing stars, with the Jhelum giants identified in this study as well as by \cite{ji20} and the potential GES stars marked in color. The Jhelum giants from \cite{ji20} are not in APOGEE; we estimate their orbital properties by using the distance modulus found by \cite{shipp18} and stress that there are large uncertainties on these calculated quantities. However, the overall trend for the Jhelum stars is consistent and follows a prograde orbit. The potential GES stars (high-$e$ with [Fe/H]$>$-1.7) show a similar trend as in the previous studies, with a narrow distribution centered on $L_{\mathrm{Z}}$=0. These metal-rich, high-$e$ stars are divided into two categories based on their APOGEE aluminum abundances: high-Al, with [Al/Fe]$>$0 (and shown in cyan), and low-Al, with [Al/Fe]$<$0 (and shown in magenta). The potential GES stars with high-Al abundances, shown as the cyan points, have lower energies and are more similar to the trend followed by the APOGEE Jhelum pointing stars as a whole, which are presumably mainly a mix Milky Way field stars. In particular, these high-metallicity, Al-enhanced (and, as seen in Figure \ref{fig:chem9}, also $\alpha$-enhanced), high-$e$ stars may be associated with Milky Way's in situ halo population, or the ``Splash'' stars \citep{belokurov20}. 

\section{Jhelum Chemistry} \label{sec:chemistry}
In this section, we present the abundance ratios for the APOGEE Jhelum giant and compare it to the chemistry found for the Jhelum red giants from \cite{ji20}. 

\subsection{$\alpha$ elements} \label{subsec:alpha}
 
The pattern of [$\alpha$/Fe] to [Fe/H] ratios is sensitive to the star formation rate in the progenitor of a galaxy -- a low mass galaxy will have fewer massive stars and hence have a knee at lower metallicity than is seen in the Milky Way's disk \citep{sheffield12}. The knee is therefore a relative chronometer for enrichment via Type II supernovae primarily (high-$\alpha$) and the onset of enrichment by Type Ia supernovae (and a corresponding shift to lower $\alpha$). 

\begin{figure}
\begin{center}
    \includegraphics[width=\textwidth,height=\textheight,keepaspectratio]{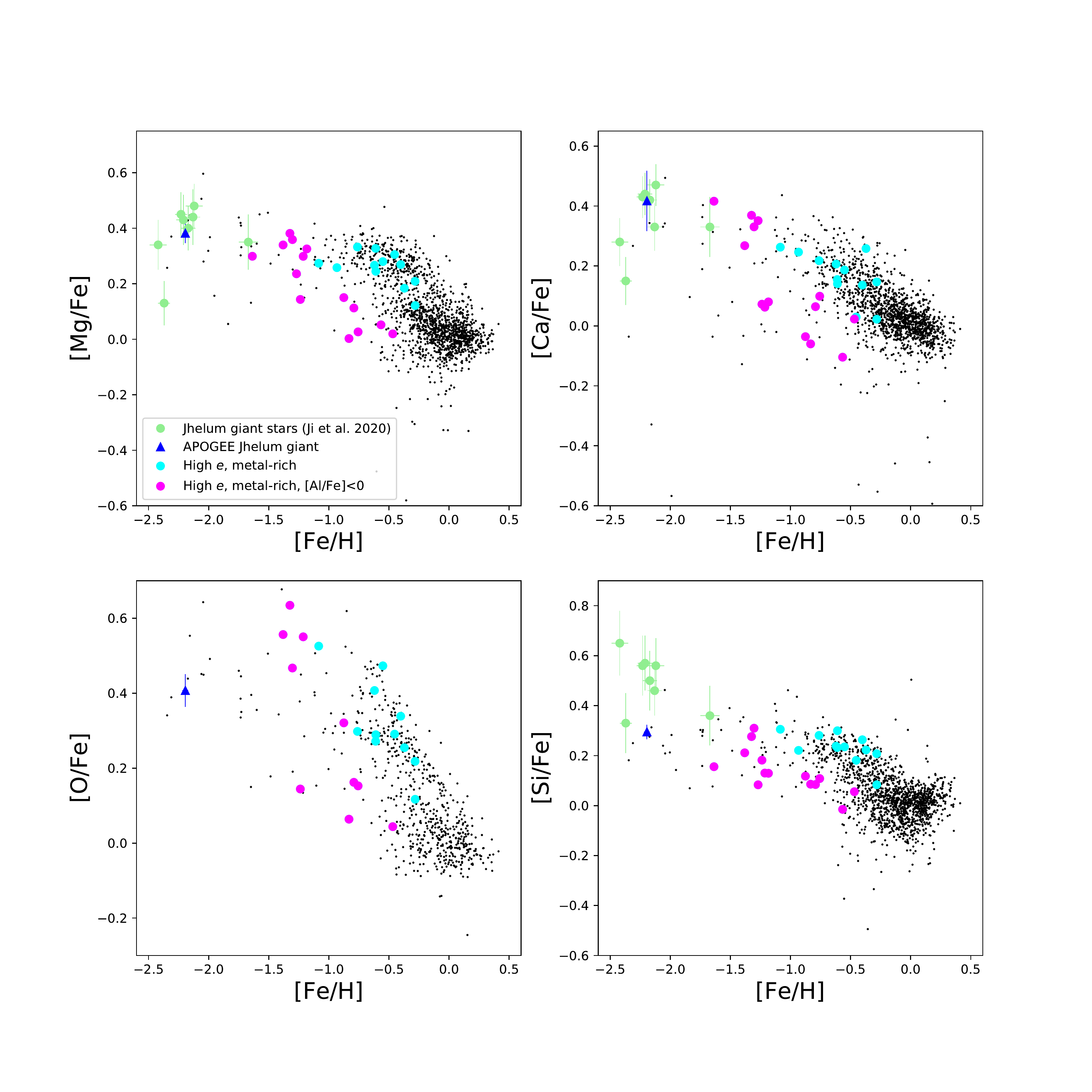}
\caption{[$\alpha$/Fe] ratios as a function of [Fe/H] for stars in the APOGEE-2 Jhelum pointings (black points). The APOGEE-2 Jhelum giant (blue triangle) from this study is shown. The Jhelum stars from \cite{ji20} are shown for \textbf{Mg, Ca, and Si} (light green circles). 
}
\label{fig:chem9}
\end{center}
\end{figure}

Figure \ref{fig:chem9} shows a summary of the $\alpha$ abundances (Mg, Ca, O and Si) as a function of [Fe/H] for the APOGEE Jhelum giant and the Jhelum giants from \cite{ji20}. The O values reported in \cite{ji20} are upper limits and we do not include them in Figure \ref{fig:chem9}. There is overlap in [Fe/H], with low scatter, seen between the APOGEE Jhelum giant and those for the \cite{ji20} Jhelum giants, with the exception of the star from \cite{ji20} at [Fe/H]=-1.67. The $\alpha$ abundances are in agreement with the general trend for Milky Way halo stars \citep[e.g., ][]{mackereth19}. The existence of high- and low-$\alpha$ components of the halo was first reported by \cite{nissen97} and subsequently found in later studies. \cite{hayes18}, for example, show that there is a division in APOGEE (DR13) between low-Mg and high-Mg stars in the halo, which in turn identifies the signature of a population of stars accreted from massive dwarf galaxies (and indeed, this population has since been argued to be dominated by a single massive accreted system, the GES remnant). The low-Mg population in \cite{hayes18} also has hot kinematics at intermediate metallicities. At lower metallicities, the low-Mg and high-Mg populations begin to converge to a constant value of around [$\alpha$/Fe] of 0.2 to 0.4, which is a value shared by both metal-poor halo stars and stars in dwarf galaxies, the likely contributors of Milky Way halo stars \citep{tolstoy09}. 
There is good agreement in [Mg/Fe] and [Ca/Fe] between the APOGEE Jhelum giant and the Jhelum giants from \cite{ji20}. However, the [Si/Fe] ratios are significantly higher for the majority of the Jhelum giants from \cite{ji20}.  This may be indicative of a genuine difference in the Si patterns, although we note that any zero-point differences between the abundance ratios derived through ASPCAP and those presented in \cite{ji20} have not been accounted for.

While APOGEE reports the measurement of other elements, many of which overlap with those presented in \citet{ji20}, many of these other elements can only be measured reliably by APOGEE at higher metallicities. Some elements, such as Na and V, only have a couple of weak lines in APOGEE's spectral coverage \citep{Jonsson_2020} and are difficult to measure in almost any metal-poor stars. Additionally, at low metallicities, other elements such as N, K, Cr or even some of APOGEE's most precisely measured elements like C, Mn, and Ni begin to weaken to the point that, depending on the abundance of that element and the star's temperature, they may not be detectable at APOGEE resolution and typical S/N depending on the abundance of that element \citep{hayes18}.  Since the APOGEE Jhelum giant is very metal-poor, APOGEE does not provide as highly reliable measurements for these elements and we have chosen to exclude them.  Finally while Al lines are often strong even for very metal-poor stars in APOGEE, we do not compare the Al of the APOGEE Jhelum giant with the results from \citet{ji20} because APOGEE's synthetic grid only extends to [Al/H] = -2.5, where the APOGEE Jhelum star is reported, so it is merely an upper limit consistent with the \citet{ji20} findings of [Al/Fe] $\approx$ -0.5 at metallicities around -2.2.  For those reasons, we have just limited our analysis here to the $\alpha$ elements which APOGEE can reliably measure even down to these low metallicities.

\subsection{Formation Scenario} \label{subsec:formation}
Stellar streams can form via the tidal disruption of a accreted galaxy or a GC.
From the \cite{ji20} study, the metallicity distribution of the Jhelum stellar stream giants, while generally metal-poor with [Fe/H] $<$ -2, span a wider region than expected for a Milky Way GC with a single stellar population. In particular, one of the \cite{ji20} Jhelum giants has a metallicity of -1.67; the presence of a metal-rich star amongst an otherwise metal-poor mono-metallic population is evidence against a GC origin scenario. We verify that this metal-rich giant follows the same orbital pattern as the more metal-poor members, including the APOGEE Jhelum giant (see Figure \ref{fig:orbits_dist}).

In \cite{bonaca2019}, the ratio of blue stragglers to blue horizontal branch stars in Jhelum stream members is found to be consistent with that of either a dwarf satellite or a low mass GC. N-body simulations showing morphological differences between a stream formed via a satellite versus GC \citep{malhan20}, in particular how the core of the dark matter subhalo impacts this morphology, show that the structure of the Jhelum stream, as well as GD-1, are consistent with a GC that formed in a satellite galaxy that the Milky Way accreted. Some of the complex features of these two streams --- such as gaps, spurs, and a less-diffuse density enhancement parallel to the main one --- can be explained by a stream that was tidally stripped while still inside its parent subhalo. Thus, globular clusters formed within accreted satellites (e.g., the GCs identified by \citealt{myeong18} that are associated with the GES merger progenitor) are a possible origin scenario.  

Chemical tagging of stars in a stream can be used to determine if a GC progenitor of a stream was born in situ or was part of a larger accreted system  \citep{horta20,meszaros20}. A Na-O anti-correlation, due to material processed during the lower temperature chain involving O-Ne-Na, is present in most GCs, although there are some notable exceptions such as Ruprecht 106 \citep{Villanova_2013}. A Si-Al correlation as well as anti-correlations for Mg-Al are found for many Milky Way GCs \citep[e.g.,][]{carretta10,Bastian_2018} and subtle but measurable differences are seen for accreted versus in situ GC stars of intermediate metallicity ($-1>$ [Fe/H] $>-1.5$) for [Si/Fe] as a function of metallicity \citep{horta20}. A trend in Mg-Al for APOGEE stars within 4 kpc of the Galactic center is also found by \cite{horta21} and is shown to reliably distinguish stars belonging to different substructures in the inner Milky Way. At the low [Al/Fe] presumed for the APOGEE Jhelum giant, as suggested by the results of \cite{ji20} (they find that all eight of their Jhelum giants have [Al/Fe]$<$0), it is difficult to test the Mg-Al and Si-Al correlations. The low aluminum abundance for the (very low metallicity) APOGEE Jhelum giant could be consistent with a first generation (primordial) GC star, a MW halo star, or a dSph star \citep{fernandezTrincado19}.

Anti-correlations between Si-Mg are also seen in GCs \citep[see, e.g., ][]{meszaros20}. \cite{hayes18} show that the high- and low-Mg populations are distinct in their [Si/Fe] ratios --- the low-Mg accreted population is Si-poor relative to the high-Mg population --- but this trend starts to blur at metallicities below $-1.5$. 
 
\begin{figure}[h]
    \centering
    \includegraphics{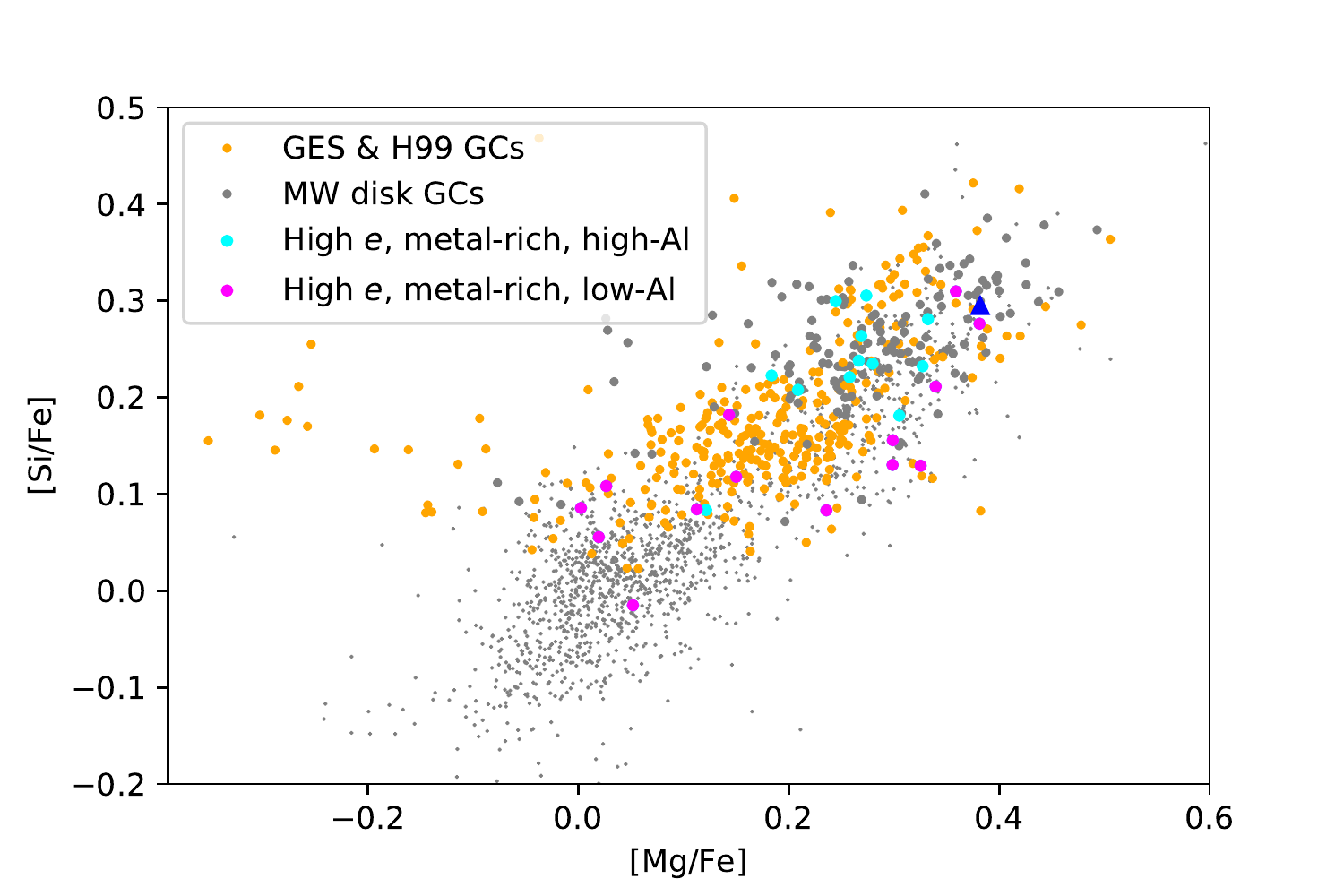}
    \caption{The [Si/Fe] versue [Mg/Fe] distributions for all Jhelum pointing stars (small grey points), the APOGEE Jhelum RGB star (blue triangle), high-$e$ metal-rich stars (cyan circles), high-$e$ metal-rich stars with [Al/Fe]$<$0 (magenta circles), accreted MW GCs (orange circles) and in situ MW disk GCs (grey circles). The accreted GCs (orange) are associated with the GES merger and the H99 streams (see \citealt{horta20} for details on the accreted/in situ classifications for the GCs).}
    \label{fig:simg}
\end{figure}

In Figure \ref{fig:simg}, we show the APOGEE Jhelum giant and the GES candidates along with both in situ and accreted GCs \citep{horta20}, along with all APOGEE DR16 stars with [Fe/H] $<$ -2.0. As seen in Figure \ref{fig:simg}, the APOGEE Jhelum giant star falls into the high-Si and high-Mg region that both accreted and in situ GCs occupy, as well as both accreted dwarf satellite stars (e.g., Sculptor and Carina) and Milky Way field stars. Thus, based on currently available abundances, we cannot narrow down the origin of the Jhelum stream based on chemistry alone.

\begin{table}
\footnotesize
\begin{center}
\begin{tabular}{llllllllll}
\hline
APOGEE ID & $T_{\rm eff}$ & log($g$) & [Fe/H] & [Mg/Fe] & [Si/Fe] & [Ca/Fe] & [O/Fe]    \\
\hline
2M22033201-4705352 & 4421 $\pm$ 103 & 0.80 $\pm$ 0.10 & -2.19 $\pm$ 0.02 & 0.38 $\pm$ 0.04 & 0.29 $\pm$ 0.03
& 0.42 $\pm$ 0.10 & 0.41 $\pm$ 0.04 \\
\hline
\end{tabular}
\caption{The ASPCAP abundance ratios for the APOGEE Jhelum giant.}
\end{center}
\end{table}

\section{Summary} \label{sec:summary}
In this study, we looked in detail at stars in the direction of the Jhelum stellar stream, taken from the APOGEE-2 database. By isolating stars that have the proper motion signal detected for Jhelum stream stars by \cite{bonaca2019} and \cite{shipp19} and then fitting this proper motion-selected sample to an isochrone for a 12 Gyr population with [Fe/H] $=-1.4$ and [$\alpha$/Fe] $=0.4$, we identified one member of the Jhelum stellar stream. This Jhelum star is a bright red giant, near the tip of the RGB for the Jhelum population.

By looking at the azimuthal and radial components of motion for stars in the six APOGEE fields targeting the Jhelum stream, we identified a group of metal-rich ([Fe/H] $>-1.7$), high eccentricity ($e>0.8$) stars that are likely to be part of the debris from the $Gaia$-Enceladus-Sausage merger. We divided this GES-sample into high-Al and low-Al groups to help distinguish stars likely belonging to merger debris, as stars from accreted populations tend to have low [Al/Fe] values. While the errors for the distances to the Jhelum giants is large, and these propagate into derived quantities like energy and angular momentum, we compare the total energy as a function of angular momentum ($L_{z}$) for the Jhelum giants to the GES giants and find that, while both populations have similar angular momenta (both are prograde), the Jhelum giants tend to have higher total energies than the GES giants. 
The orbit for the APOGEE Jhelum giant and the APOGEE GES giants were integrated forward for 5 Gyr and compared, and their respective orbits do not seem consistent with stars derived from same population. In particular, the orbits for the GES giants are highly radial and tend to stay more confined to the plane than does Jhelum.

Unfortunately, at the low metallicities of the Jhelum giants, the $\alpha$-element abundance patterns (for the elements Mg, Ca, O, and Si) for the Jhelum stellar stream overlap with the locations of both accreted satellite stars as well as stars from both accreted and in situ globular clusters, and thus we cannot distinguish between these scenarios. Recent work by \citet{bonaca20} lends support to the accreted galaxy origin for the progenitor of the Jhelum stream, and find tentative evidence that either the H99 streams \citep{helmi99} or the Wukong structure \citep{naidu20} are potential candidates for the Jhelum progenitor.

\acknowledgments

We thank the SDSS Faculty And Student Teams (FAST) Initiative and in particular Jesus Pando, for providing support for this work, in particular the CUNY FAST group. 

Support for this work was provided by NASA through Hubble Fellowship grant \#51386.01 awarded to R.L.B. by the Space Telescope Science Institute, which is operated by the Association of  Universities for Research in Astronomy, Inc., for NASA, under contract NAS 5-26555. SRM acknowledges support from National Science Foundation grants AST-1616636
and AST-1909497. DAGH acknowledges support from the State Research Agency (AEI) of the Spanish Ministry of Science, Innovation and Universities (MCIU) and the European Regional Development Fund (FEDER) under grant AYA2017-88254-P.

Funding for the Sloan Digital Sky Survey IV has been provided by the Alfred P. Sloan Foundation, the U.S. Department of Energy Office of Science, and the Participating Institutions. 

SDSS-IV acknowledges support and resources from the Center for High Performance Computing  at the University of Utah. 
The SDSS website is \url{www.sdss.org}.

SDSS-IV is managed by the Astrophysical Research Consortium for the Participating Institutions of the SDSS Collaboration including the Brazilian Participation Group, the Carnegie Institution for Science, Carnegie Mellon University, Center for Astrophysics | Harvard \& Smithsonian, the Chilean Participation Group, the French Participation Group, Instituto de Astrof\'isica de Canarias, The Johns Hopkins University, Kavli Institute for the Physics and Mathematics of the Universe (IPMU) / University of Tokyo, the Korean Participation Group, Lawrence Berkeley National Laboratory, Leibniz Institut f\"ur Astrophysik Potsdam (AIP),  Max-Planck-Institut f\"ur Astronomie (MPIA Heidelberg), Max-Planck-Institut f\"ur Astrophysik (MPA Garching), Max-Planck-Institut f\"ur Extraterrestrische Physik (MPE), National Astronomical Observatories of China, New Mexico State University, New York University, University of Notre Dame, Observat\'ario Nacional / MCTI, The Ohio State University, Pennsylvania State University, Shanghai Astronomical Observatory, United Kingdom Participation Group, Universidad Nacional Aut\'onoma de M\'exico, University of Arizona, University of Colorado Boulder, University of Oxford, University of Portsmouth, University of Utah, University of Virginia, University of Washington, University of Wisconsin, Vanderbilt University, and Yale University.

\vspace{5mm}

\software{  Astropy \citep{Astropy_2013,Astropy_2018}, 
            NumPy \citep{vanderWalt_2011,Harris_2020numpy}, 
            gala \citep{gala:zenodo,gala}
            }

\facilities{Du Pont (APOGEE), Sloan (APOGEE), \emph{Gaia}}

\bibliography{sample63.bib,apogee2_refs.bib}{}
\bibliographystyle{aasjournal}


\end{document}